\documentclass{aipproc}
\usepackage{natbib}
\layoutstyle{6s}
\def\msun{\mbox{${M_\odot}$}}
\def\mpc{\mbox{$\;{\rm Mpc}$}}

\def\kms{\mbox{$\;{\rm km\ s}^{-1}$}}     
\def\mev{\mbox{$\;{\rm MeV}$}}
\def\kev{\mbox{$\;{\rm keV}$}}
\def\gev{\mbox{$\;{\rm GeV}$}}
\def\gyr{\mbox{$\;{\rm Gyr}$}}
\def\ghz{\mbox{$\;{\rm GHz}$}}
\def\jy{\mbox{$\;{\rm Jy}$}}

\def\cc{\mbox{$\;{\rm cm}^{-3}$}}
\def\mug{\mbox{$\;\mu$G}}
\def\lesssim{\lower.5ex\hbox{$\; \buildrel < \over\sim \;$}}

\def\ergs{\mbox{$\;{\rm ergs}$}}
\def\ergss{\mbox{$\;{\rm ergs\ s^{-1}}$}}

\def\etal{{\it {et al.}}}
\def\aap{{\it {Astron. \& Astrophys.}}}

\def\apj{{\it {ApJ}}}
\def\apjl{{\it {ApJL}}}

\def\asph{{\it {Astroparticle Physics}}}

\def\mnras{{\it {MNRAS}}}
\def\nat{{\it {Nature}}}
\def\newa{{\it {New Astronomy}}}
\def\pasp{{\it {Publ. Astr. Soc. Pacif.}}}

\begin{document}
\title{Accelerated Particles from Shocks Formed in Merging Clusters of
Galaxies}

\author{Robert C.\ Berrington}{
address={ASEE Postdoctoral Fellow, 
Naval Research Laboratory, Code 7653, Washington, DC 20375-5352},
email={rberring@gamma.nrl.navy.mil}
}

\author{Charles D.\ Dermer}{ address={Naval Research Laboratory, Code 7653,
Washington, DC 20375-5352},email={dermer@gamma.nrl.navy.mil} }

\author{S.\ J.\ Sturner}{ address={NASA's GSFC and
USRA, 7501 Forbes Blvd. \#206, Seabrook, MD
20706-2253},email={sturner@swati.gsfc.nasa.gov} }

\begin{abstract}
Subcluster interactions within clusters of galaxies produce shocks that
accelerate nonthermal particles. We treat Fermi acceleration of nonthermal
electrons and protons by injecting power-law distributions of particles during
the merger event, subject to constraints on maximum particle energies. The
broadband nonthermal spectrum emitted by accelerated electrons and protons is
calculated during and following the subcluster interaction for a standard
parameter set.  The intensity of $\gamma$-ray emission from primary and
secondary processes is calculated and discussed in light of detection
capabilities at radio and $\gamma$-ray energies.
\end{abstract}

\maketitle

\section{Introduction}

Rich clusters contain thousands of galaxies and are the largest
gravitationally bound systems in nature.  Masses for rich clusters are
$\sim\!\!10^{15} \msun$, with $\sim \!\!5$--10\% of the mass found in a hot
intergalactic gas at temperatures of 2--12\kev.  Rich clusters emit thermal
bremsstrahlung with luminosities $L_{x} \sim 10^{44}-10^{45}$\ergss
\citep{allen:94}. Poor clusters contain hundreds of galaxies, have total
masses $\sim\!\!10^{14} \msun$, have a hot intergalactic gas of temperatures
1--5\kev\ and X-ray luminosities $L_{x} \sim 10^{42}-10^{43}$\ergss
\citep{dahlem:00}. Approximately 90\% of the total mass of clusters is in the
form of nonluminous dark matter.

In the hierarchical merging cluster scenario, poor clusters merge together to
form richer clusters.  Approximately 30--40\% of galaxy clusters show evidence
of substructure in both the optical \citep{beers:82} and X-ray wavelengths
\citep{forman:81}.  Velocity differences between the observed structures is
$\approx\!\!1000$--$2000$\kms.  With gravitational forces driving the
interaction between the two systems, cluster mergers are consistent with
highly-parabolic orbits.  Typical sound speeds within the intergalactic medium
(IGM) are $\approx\!\!800$\kms, so shocks will form at the interaction
boundary of the two systems.  Computer models of merging clusters support the
development of shocks in the IGM
\citep{roettiger:93,ricker:98,takizawa:99}. Dimensional arguments show that a
cluster merger releases $10^{63}-10^{64}$\ergs\ of gravitational potential
energy when initial separations are of the order $\sim\!\!$ Mpc.

Only the most massive and X-ray luminous galaxy clusters have extended diffuse
radio sources.  With projected linear sizes $\sim\!\!1$\mpc, these diffuse
sources have no known optical counterparts.  The diffuse radio emissions have
two distinct characteristics.  The extended diffuse emission found in the
central region of a galaxy cluster with a regular, azimuthally symmetric shape
is known as a {\em radio halo}. The diffuse emission found on the cluster
periphery are the cluster {\em radio relics}.  These features often have
irregular shapes with signs of filamentary structure.  Radio relics are
associated only with clusters that show evidence of a recent or ongoing merger
event.

Shock fronts that form in the IGM as a result of a cluster merger event are
thought to be associated with the cluster radio relics.  The shock compression
will orient any existing cluster magnetic field into the plane of the shock.
Radio relics are characterized by highly organized magnetic fields with field
strengths in the $\sim\!\!1$\mug\ range with linearly polarized field lines in
the vicinity of the shock \citep{ensslin:98}.  The shock front will accelerate
a fraction of the thermal particles within the IGM by first-order Fermi
acceleration.

Recent work has highlighted the importance of nonthermal radiation from
particles accelerated by shocks formed in merging clusters. Loeb and Waxman
\citep{loeb:00} have proposed that cluster mergers are the dominant
contributor to the diffuse $\gamma$-ray background, provided the efficiency to
convert the available gravitational energy into nonthermal electron energy is
$\sim \!\!5$\%.  Some unidentified EGRET sources are claimed to be associated
with $\gamma$-ray emission from galaxy clusters \citep{colafrancesco:98}.
Excess EUV emission from Coma can be explained by nonthermal electrons
accelerated at merger shocks \citep{atoyan:00}.  Variations in radio surface
brightness will result from superposition of cluster emissions
\citep{waxman:00}.

In order to examine this question in more detail, we have modified a supernova
remnant code \citep{sturner:97} to calculate nonthermal radiation spectra from
primary and secondary particles in merger shocks. We go beyond previous
treatments by considering both primary electron and proton acceleration, a
time-dependent treatment of radiation losses, and radiation signatures of
secondaries from proton-nuclear interactions.

\section{Model}

We have adapted a supernova remnant code \citep{sturner:97} to treat the
cluster merger scenario.  The code is designed to calculate time-dependent
particle distribution functions evolving through adiabatic and radiative
losses for electrons and protons accelerated by the first-order Fermi process
at the cluster merger shock.

The electron and proton distribution functions originate from a momentum
power-law injection spectrum. In terms of total particle energy $E = m\gamma
c^2$, the injection function is
\begin{equation}
Q_{e,p}(E,t) = Q_{e,p}^{0} \left[\frac{(pc)^{-s}}{\beta} \right] \exp
\left[ - \frac{E}{E_{max}(t)} \right],
\label{eqn:power-law}
\end{equation}
where $p=\beta\gamma$ is the dimensionless momentum and $s$ is the injection
index.  The maximum particle energy $E_{max}$ is determined by the maximum
energy associated with the available time since the beginning of the merger
event, by a comparison of the Larmor radius with the size scale of the system,
and by a comparison of the energy-gain rate through first-order Fermi
acceleration with the energy-loss rate due to adiabatic, synchrotron and
Compton processes.  Particle injection ceases after the age of the shock front
exceeds $ t_{acc} = 10^9$ yrs. The constant $Q_{e,p}^{0}$ normalizes the
injected particle spectrum over the entire volume $V(t)$ swept out by the
shock front, and is determined by
\begin{equation}
E_{e,p}^{tot} = \int_{0}^{t_{acc}}\ dt\ \int_{0}^{E_{max}}\ dE\ E\
Q_{e,p}(E,t)\ V(t)\;.
\label{eqn:normalization}
\end{equation}
We assume a total available energy $E_{e,p}^{tot} =\eta_{e,p} 10^{63}$\ergs,
and an efficiency factor $\eta_{e,p} = 5\%$ for both protons and
electrons. Although the injection index $s$ depends upon the Mach number of
the shock, here we present calculations for a fixed index $s=2$.

The time evolving particle spectrum is determined by solving the Fokker-Planck
equation in energy space for a spatially homogeneous IGM, given by
\begin{equation}
\frac{\partial n(E,t)}{\partial t} = - \frac{\partial}{\partial
E}[\dot{E}_{tot}(E,t) n(E,t)] + \frac{1}{2} \frac{\partial^{2}}{\partial
E^{2}}[D(E,t) n(E,t)] + Q(E,t) - \frac{n(E,t)}{\tau_{pion}(E,t)}\;.
\label{eqn:Fokker-Planck}
\end{equation}
The quantity $\dot{E}_{tot}(E,t)$ represents the total synchrotron, Compton,
Coulomb, and adiabatic energy-loss rate for electrons, and the sum of the
Coulomb and adiabatic energy-loss rates for protons.  Both protons and
electrons are subject to diffusion in energy space by Coulomb interactions.
The protons experience catastrophic losses due to proton-proton collisions on
the time scale $\tau_{pion}$.  The spectra of secondary electrons and
positrons are calculated from pion-decay products, and are subject to the same
physical processes as the primary electrons.

The synchrotron, Compton, bremsstrahlung, and pion-decay $\gamma$-ray spectral
components are calculated from the particle spectra following the methods
described by \citet{sturner:97}.  We use a standard parameter set with a mean
IGM number density $n_{IGM} = 10^{-3}\cc$, a uniform cluster magnetic field of
$B = 0.1\mug$, a constant shock speed $v_s = 1000$\kms, and an acceleration
period of 1\gyr. Thus, $V(t) = A(t) v_s t$.  We assume a constant surface area
with a 1 Mpc radius. Particles loose energy due to adiabatic expansion
according to the relation $-\dot\gamma/\gamma = \dot V(t)/V(t)= 1/t$.

\begin{figure}[t]
\includegraphics[width=6in]{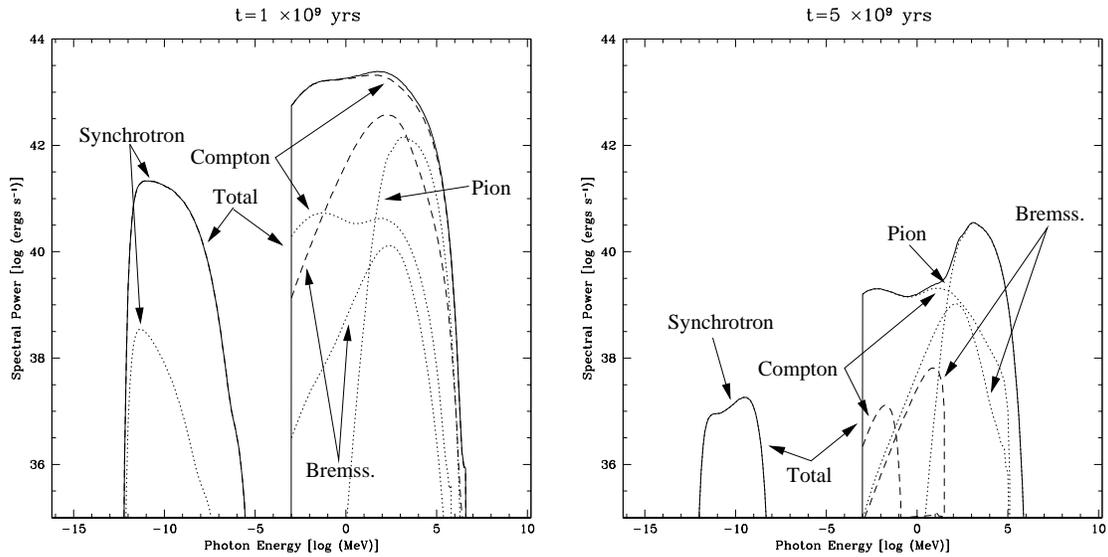}
\caption{Total nonthermal photon energy spectra from a cluster merger shock.
The solid curves are the total photon spectra summed from all processes.  The
dashed curves are the photons produced from the primary electrons, and the
dotted curves are the secondary emission components.  The left and right
panels show the total nonthermal photon spectra at $t=1$\gyr\ and $t=5$\gyr,
respectively, after the onset of the cluster merger event.}
\label{fig:15gyr}
\end{figure}

\section{Results and Discussion}

Figure \ref{fig:15gyr} shows nonthermal photon spectra calculated at 1\gyr\
and 5\gyr\ for our standard parameters.  The system is very luminous at radio
frequencies during the particle acceleration phase because of intense Compton
losses of nonthermal electrons on the CMB. After the acceleration period ends,
the radiation from primary electrons is dominated by emission from
secondaries.  The $\pi^0$ bump at 70 MeV is hidden beneath the
Compton-scattered CMB radiation from primary electrons during the acceleration
period, but dominates the $\gamma$-ray spectrum after the acceleration period
is over. The intensity of the $\pi^0$ bump depends sensitively upon the
relative efficiencies $\eta_{p,e}$ for proton and electron acceleration. If no
pion signature is found in nonthermal $\gamma$ rays from merging clusters,
then $\eta_p\lesssim \eta_e$ for our standard parameters.

\begin{figure}
\includegraphics[width=6in]{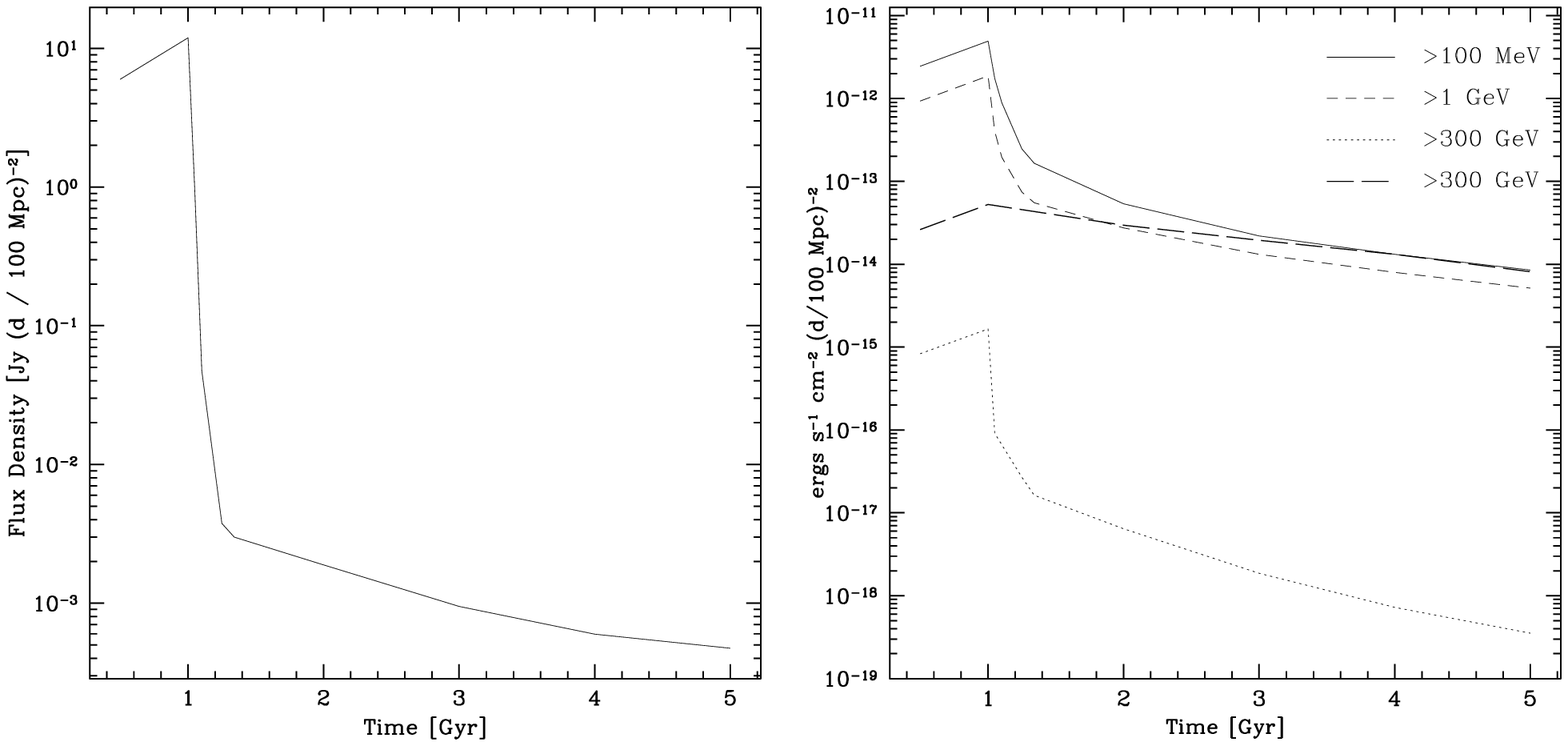}
\caption{Calculated 1.4\ghz\ radio flux density (left panel) and integrated
flux of high energy photons (right panel) from a cluster merger shock.  Both
panels are normalized for a system at a distance of 100\mpc.  The radio flux
density is in the 1.4\ghz\ band, and the integrated energy fluxes for the high
energy photons are shown in the right panel. Also shown is the $>\!\!300$ GeV
emission for a merger where $v_s = 2000$ km s$^{-1}$ and $B = 1~\mu$G as the
boldfaced, long-dashed line.}
\label{fig:flux}
\end{figure}

From the total photon spectra in Figure \ref{fig:15gyr}, we have calculated
light curves at various photon energies in Figure \ref{fig:flux} for a merging
cluster at a distance of 100\mpc.  The maximum peak 1.4 GHz radio flux density
of $\sim\!\!$ 12\jy\ occurs at the end of the acceleration period.  The radio
emission from the merger event is, however, distributed over an angular region
$\sim\!0.5^\circ$ of the merger shock, making it difficult to detect.

Figure \ref{fig:flux} presents calculations of the energy fluxes from a merger
shock at 100 Mpc, for photon energies $>\!\!100$\mev, $>\!\!1$\gev, and
$>\!\!300$\gev.  The maximum luminosity occurs at the end of the acceleration
period due to the accumulation of nonthermal protons.  Cooling times due to
synchrotron and Compton energy losses are very short for high-energy
electrons; electrons that Compton-scatter microwave photons to energies $E =
100 E_{100}$ MeV will cool on a time scale of $t_{cool} \cong 6\times
10^6/\sqrt{E_{100}}$ yrs. Consequently, the $\gamma$-ray emission declines
sharply after particle acceleration ceases, and approaches a level where all
the $\gamma$-ray production originates from proton interactions. The maximum
flux for secondary electrons is $\approx\!2$ orders of magnitude less than the
maximum flux for the primary electrons, as can be seen by comparing the
catastrophic proton loss time scale of $\sim\! 30$\gyr\ with the 1 Gyr
injection time scale.  Because of adiabatic losses and the long catastrophic
pion production cooling times, protons inject a slowly declining rate of
secondary electrons in the 1-4 Gyrs after the initial acceleration event.

\section{Summary}

We have studied the acceleration of protons and electrons in shocks formed by
merging clusters of galaxies, and calculated the expected synchrotron,
Compton, bremsstrahlung, and pion emission from the accelerated particles.
Both the radio flux density at 1.4\ghz\ and the $\gamma$-ray flux at photon
energies $>\!\!100$\mev, $>\!\!1$\gev\ and $>\!\!300$\gev\ for a source at
100\mpc\ as a function of time are presented.  The flux is greatest at the end
of the particle acceleration phase, and merging clusters of galaxies at a
characteristic distance of 100\mpc\ should be detectable with radio telescopes
and GLAST.  Radio and $\gamma$-ray detectability of merger events must contend
with the angular extent of the emission.

Because the merger event lasts for only $\sim\!\!1$ Gyr, a large fraction of
clusters will no longer be experiencing ongoing nonthermal particle
acceleration from cluster mergers. GLAST, with a limiting sensitivity of
$4\times 10^{-13}$ ergs cm$^{-2}$ s$^{-1}$ for a one year all-sky survey, will
still be able to detect such systems if they are at optimal distances.
Depending on the shock speed and magnetic field, cluster merger events can
also be detectable sources of TeV radiation (see Fig.\ 2).  We therefore
expect detectable $\gamma$-ray emission from clusters at distances $\sim\!\!
100$\mpc. Within this distance a number of radio relics have been detected
\citep{giovannini:00,kempner:01}.

\end{document}